\documentclass[prl,twocolumn,showpacs,amsmath,amssymb]{revtex4}
\usepackage{graphicx}
\usepackage{dcolumn}
\usepackage{bm}

\begin{document}


\title{Local electronic properties in the superconducting and the normal phase in the disordered film of titanium nitride}
\author{P. Kulkarni$^{*}$}
\affiliation{Laboratorio de Bajas Temperaturas, Departamento de F\'isica de la Materia Condensada, Instituto de Ciencia de Materiales Nicol\'as Cabrera, Facultad de Ciencias, Universidad Aut\'onoma de Madrid, 28049 Madrid, Spain}
\author{H. Suderow}
\affiliation{Laboratorio de Bajas Temperaturas, Departamento de F\'isica de la Materia Condensada, Instituto de Ciencia de Materiales Nicol\'as Cabrera, Facultad de Ciencias, Universidad Aut\'onoma de Madrid, 28049 Madrid, Spain}
\author{S. Vieira}
\affiliation{Laboratorio de Bajas Temperaturas, Departamento de F\'isica de la Materia Condensada, Instituto de Ciencia de Materiales Nicol\'as Cabrera, Facultad de Ciencias, Universidad Aut\'onoma de Madrid, 28049 Madrid, Spain}
\author{M.R. Baklanov}
\affiliation{IMEC, Kapeldreef 75, B-3001 Leuven, Belgium}
\author {T. Baturina}
\affiliation{Institute of Semiconductor Physics, 13 Lavrentjev Avenue, Novosibirsk, 630090, Russia}
\author{V. Vinokur}
\affiliation{Materials Science Division, Argonne National Laboratory, Argonne, Illinois 60439, USA}

\date{\today}

\begin{abstract}
We present in this paper the conductance maps at 100 mK in the disordered polycrystalline film of titanium nitride (TiN). At 5 nm, the film is close to quasi-two dimensional limit and exhibits features pertaining to the superconductor to insulator transition\cite{Baturina,Baturina1}. We measured conductance maps at zero field and at 4 T, which represent the superconducting and the normal phase, respectively. The conductance map at 4 T is uniform, in which the conductance behavior, with logarithmic variation, resembles to the disorder enhanced electron-electron interaction in the two dimensional metallic phase. At low fields we observe the spatial variations of the conductance in the superconducting phase. At several places the superconducting energy gap fluctuates to an extent that the quasi-particle peaks are absent in the conductance curves. The conductance map over a region encompassing only few crystallites suggests that the inhomogeneities in the superconducting phase related to the spatial variations of the electronic density are across the crystalline boundaries.
\end{abstract}


\maketitle

It is well known that temperature, magnetic field and critical current density destroy the superconducting behavior, but the changes introduced by disorder or reduction of sample dimensions are not so well known. Strongin el al.\cite{Strongin70} showed that the critical temperature $T_c$ of continuous superconducting films decreases when its thickness is reduced close to few atomic layers and a subsequent transition was observed towards an insulating ground state. Now it is understood that disorder plays a key role in this transition\cite{Dubi,Feigelman,Fisher,Haviland}. In thin superconducting films, disorder drives the film from a superconductor to a "bad metal" behavior at finite temperatures which strongly resembles an insulator. The transition occurs typically when the resistance crosses the quantum of resistance $h/(e)^2$ by varying thickness or magnetic field\cite{Fisher,Haviland,Hebard,Sambandamurthy,Sambandamurthy1,Bielejec,Bielejec1}. For different levels of disorder, either superconductivity or insulating behavior can follow and the resistance vs temperature curve displays preceeding negative slope\cite{Baturina,Baturina1}. A high field behavior with relatively low resistance is found in both superconducting and insulating samples close to the transition\cite{Baturina}. Previous zero field Scanning Tunneling Microscopy (STM) work shows spatial fluctuations in the superconducting properties at distances of the coherence lenght InO and TiN thin films\cite{Sacepe,Sacepe1}. Local variations in both the superconducting gap and the height of the coherence peaks were reported in line scans and sets of a hundreds of tunneling conductance curves, pointing out that the superconducting properties are inhomogeneous. Temperature dependence gives the disappearance of the superconducting quasiparticle peaks at the onset of resistance and the appearance of a pseudogap like background surviving up to temperatures several times T$_c$\cite{Sacepe,Sacepe2}. New concepts have been tested, such as Cooper pair localization at the onset of SIT, leading to the so-called Bose-Insulator, or the destruction of Cooper pairs from Coulomb interaction, leading to the Fermi insulator\cite{Dubi,Feigelman,Galitski,Gantmakher,Butko,Bouadim}. These theories raise rather basic questions about the transition: What is the role of sample and its microscopics? Are sample inhomogenieties at nanoscales contribute to changes in the superconducting phase? Further, do the normal phase shows the electronic behavior related to the presence of disorder? Here we present new very low temperature Scanning Tunneling Microscopy and Spectroscopy (STM) with full high resolution large conductance maps, involving tens of thousands of tunneling conductance curves and study the magnetic field dependence. Our study gives some insight into the local electronic behavior in the superconducting and the normal phase in the disordered film of TiN.

\begin{figure}[h]
\includegraphics[width=0.48\textwidth]{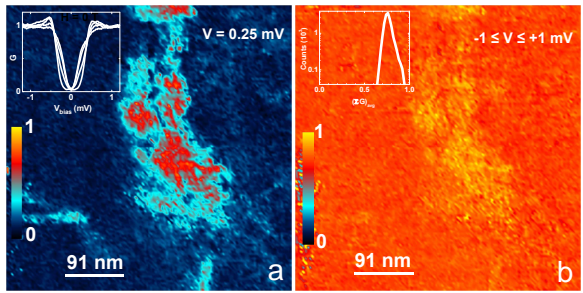}
\caption{Main panel (a) shows the conductance distribution at V$_{bias}$ = 0.25 mV spread over an area 455 nm x 455 nm measured at zero magnetic field. The color scale represents the normalized conductance with dark blue regions in the map have near zero values and the red colored regions showing conductance upto 0.75, i.e, 75 percent of the normalized conductance. The representative conductance curves are shown in the inset in panel (a). The curve displaying the clear quasi-particle peaks is measured at dark blue regions in the main panel (a), and the curve with monotonic decrease of conductance as the bias voltage approaches zero mV is the representative of the red regions in the map. Two intermediate curves measured in the vicinity of these two regions are also shown in the inset. In panel (b) the color scale represents the average of the sum of the conductance for each curve, in the bias voltage range from -1 to +1 mV. In the inset in panel (b), the histogram of the average of the sum of the conductance, with 128 x 128 values, is shown and their spatial map is shown in the main panel (b).}
\end{figure}

We study a disordered polycrystalline TiN thin film with a thickness of about 5 nm and $k_F \ell$ = 1.5, obtained from resistance. The crystallites range from 3 to 10 nm, with mean size around 5 nm. This film has a superconducting transition at $\approx$1 K and is marginally close to the superconductor to insulator transition (sample D15 of Ref.\cite{Pfuner}). Note however that the sample shows an increase of the resistance with cooling, accompanied by a slight decrease of the resistance at some hundred mK. Scanning tunneling microscopy and spectroscopy were carried out using a home made STM set up in a dilution refrigerator which cools down to 100 mK. Magnetic field was applied perpendicular to the surface of the sample. We take full (128$\times$128) conductance maps at 100 mK, over a preselected area after normalizing the tunneling conductance curves. We show here those particular bias voltages where constrast in the tunneling conductance is the highest (0.25 mV), and also give the full bias voltage dependence in a separate figure. 

\begin{figure}[h]
\includegraphics[width=0.48\textwidth]{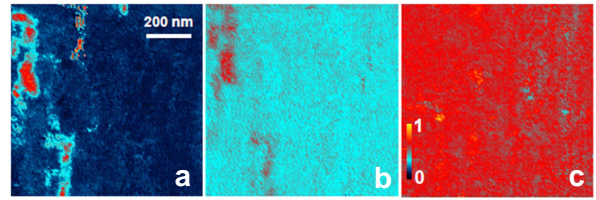}
\caption{In the main panel (a) the white color bar at the top right corner is the length scale of the area 950 nm x 950 nm. The panel (a) is the conductance map at $V_{bias}$ = 0.25 mV at H = 0 T. The color changes correspond to the conductance values following the scale shown in the panel (c), which shows the map at H = 4 T. Total 128 x 128 conductance values are plotted over a normalized scale of 0 to 1 and it is also common to the conductance map at H = 1.5 T shown in panel (b).}
\end{figure}

The tunneling conductance behavior changes spatially as shown in the inset in the panel Fig.1 (a). The area under investigation is 455 nm x 455 nm and for the most part of it the superconducting gap opens with the BCS-like quasi-particle peaks, however, in several curves the peaks are either suppressed or altogether absent, with the equivalent rounding near zero bias voltage. In order to generate a picture of the conductance distribution at the selected bias voltage of 0.25 mV, all the conductance curves are normalized at 1 mV. The normalized conductance varies from 0 to 1, as shown in the color scale in Fig. 1(a). The dark blue color is the region where the conductance at 0.25 mV reaches near zero values, while the red region has the conductance close to 0.7 and is also a representative of the region where the curves without quasi-particle peaks are seen. Spatial changes were reported previously in terms of the average superconducting gap of Ref.\cite{Sacepe1}, however a clear distinction between two behaviors was not established. In the panel (b), the average of the sum of the conductance in the bias voltage range -1 to +1 mV (on the color scale from 0 to 1) for each of the 128 x 128 curves is plotted over the same area. The average of the sum of the conductance values in the form of the histogram are shown in the inset in panel (b). The histograms peaks at 0.75, and the average conductance spatially varies in a range about 30 percent to that of the full range. In particular 5 percent change at the right edge of the histogram pertains to the slight contrast present between the regions with two distinct conductance behavior shown in panel (a). The other observation from panel (b) is that the average conductance deviates (in fact, is lower) from unity which it would be in the case of a BCS-superconducting curve. 

In figure 2 we show the conductance maps at three selected magnetic fields, H = 0 T, 1.5 T and 4 T, respectively in the panels (a-c). The conductance maps are plotted at a bias voltage 0.25 mV, with the curves having been normalized at 1 mV. The area under the investigation is 950 nm x 950 nm, and is the same at three selected fields, except for a slight shift of the order of few nanometers at higher fields. The color scale in the panel (c) is also the same as that in the panel (a) in Figure 1. The regions in the red color in the panel Fig. 2(a) are those with the conductance curves without quasi-particle peaks at zero magnetic field. The contrast between two regions reduces with the increase in the magnetic field to 1.5 T. The conductance map at 4 T does not show any contrast between the two regions, instead near uniform conductace values are spread over the region where originally a strong change in the contrast was observed at 0 T. At the outset such dramatic change in the conductance maps with application of the magnetic field asserts that the inhomogeneinities at 0 T are resulting from the local electronic properties of the superconducting phase.

Further insight can be obtained from the representative conductance curves in both the regions at the selected magnetic fields, as shown in Fig. 3. In panel (a), the curves with quasi-particle peaks shown in blue color are at zero magnetic field, and with the increase in the field to 1.5 T, the peaks are suppressed, and disappear at 4 T. The change in the conductance was highlighted in the conductance maps in Fig. 2 using the color scale from 0 to 1 at 0.25 mV. In the regions where quasiparticle peaks are absent, the conductance curves merely widen at higher fields upto 4 T, and the near uniform conductance is seen even in the rest of the film for all bias voltages. The conductance curve at 4 T also uniformly shows a considerable suppression near zero bias voltage in this film. In order to compare the changes at zero magnetic field, with respect to the conductance at 4 T, the normalized curves are shown in the panels, (c) and (d), respectively for the representative curves in both the regions. With this procedure, the conductance curve at 4 T is brought to unity at all the bias voltage values, and the relative differences are seen as the increased quasi-particle peaks, particularly in the panel (c) for the conductance curve at 0 T. The panel (d) also shows quasi-particle peaks at 0 T (i.e., the increase in the conductance with respect to the condutance at 4 T), however, considerably suppressed compared to those in panel (c). At 1.5 T, the quasi-particle peak heights are similar for both the regions. This observation points out that on the background of the nearly equal average conductance (i.e., at 4 T), the quasi-particle peaks develop with different heights, while forming the superconducting phase in this film. It also directly connects to the spatial map of the average conductance in Fig. 1(b), showing a slightly higher contrast in the central region, and a sharper edge in the histogram peak shown in its inset. 

\begin{figure}[h]
\includegraphics[width=0.48\textwidth]{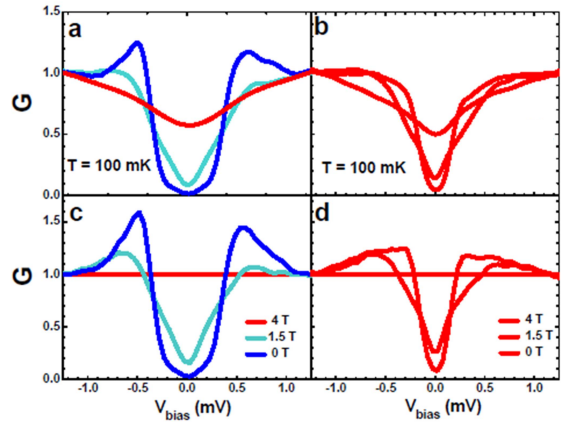}
\caption{\label{fig1} In panel (a) and (b) we show the conductance curves at selected fields in dark blue and red color regions, respectively, in Fig. 2(a), i.e., both the regions. In panels (c) and (d) we plot the conductance at 4 T as 1 at all bias voltages and renormalize the conductance at 0 and 1.5 T with respect to the 4 T curve.}
\end{figure}

In Fig. 4, we provide the conductance maps at selected bias voltages from 0.0 mV to 0.75 mV at an interval of 0.05 mV. In each panel, the bias voltage is shown in the top left corner, and the color scale represents the change of the conductance pertaining to the maximal values at that bias voltage. Such a represenation is chosen in order to highlight visually the change in the spatial conductance maps with the increase in the bias voltage. At V = 0.0 mV, a small contrast is seen in the region which have the conductance curves without the quasi-particle peaks. The contrast between the two regions increases as the bias voltage is increased, and at 0.25 mV, the maximum contrast can be seen in the conductance map. With further increase in the bias voltage, the contrast decreases, and nearly disappears at 0.4 mV, except at the interface between the two regions, where a white contrast start to appear which corresponds to the lower conductance values for these curves. The uniform contrast at 0.4 mV corresponds to the crossover of the majority of the tunneling conductance curves. From 0.45 mV the bright regions highlight the reversal of the relative conductance changes between the two regions, and it continues to peak at the bias voltage of 0.55 mV, also the position of the quasi-particle peaks in the BCS-like curves. Further increase in the bias voltage decreases the contrast, and near uniform conductance maps are seen from 0.7 mV onwards upto 1 mV. The uniform conductance (without any trace of two regions) at 0.75 mV suggests that the spatial conductance changes occur only in the range of bias voltage, corresponding to the opening of the gap, asserting the locally varying electronic behavior in the film with the formation of the superconducting phase. 

\begin{figure}[h]
\includegraphics[width=0.48\textwidth]{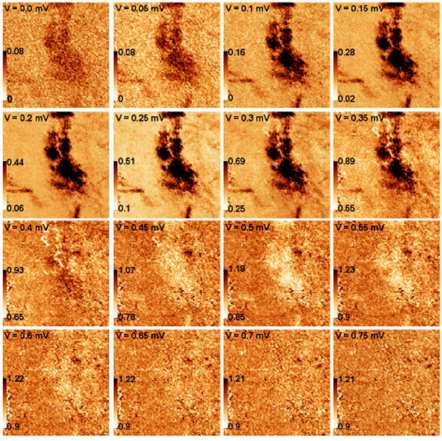}
\caption{\label{fig1} The conductance maps made at 0.05 mV interval in the bias voltage range from 0 to 0.75 mV. The labels in the top left corner in each panel show the bias voltage for each conductance map. The scale in the bottom left corner shows the range of the conductance which is kept variable in order to show the maps with the same contrast.}
\end{figure}

The main panel in fig. 5 shows the conductance behavior at 4 T, in the bias voltage range from -60 mV to +60 mV, at T = 0.15 K. The tunneling conductance monotonically reduces as zero bias votage is approached, and the peculiar variation on the logarithmic scale, shown in the inset, has perfectly linear behavior. In Fig. 6 we show at the selected magnetic fields between 0 T to 5 T, the zero bias conductance in the curves at 100 mK. The initial increase in the zero bias conductance is gradual, whereas a saturation behavior is observed above 2.5 T. The slope of this curve is plotted in the inset, and a peak in the slope is seen, marked as $H_{max}$, at 2.5 T. This value is very close to the estimated $B_{c2}$ from the resistance measurements in this film suggesting that the conductance at 4 T, and onwards corresponds to the normal phase of the TiN film. This also highlights that the conductance curves at 4 T, and subsequently the conductance maps, do not evolve with the increase in the magnetic field to 5 T. Another significant observation that the variation of the conductance is logarithmic directly relates to the Coulomb electron-electron repulsion being enhanced in the quasi-2D films of TiN due to the presence of disorder. 

\begin{figure}[h]
\includegraphics[width=0.48\textwidth]{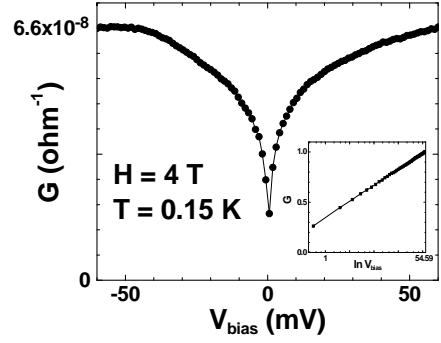}
\caption{\label{fig1} In the main panel the conductance curve at T = 150 mK is shown in the bais voltage range from -60 to + 60 mV. The curve corresponds to the magnetic field H = 4 T, and in the inset the same curve is shown in the logarithmic scale of the bias voltage.}
\end{figure}

\begin{figure}[h]
\includegraphics[width=0.48\textwidth]{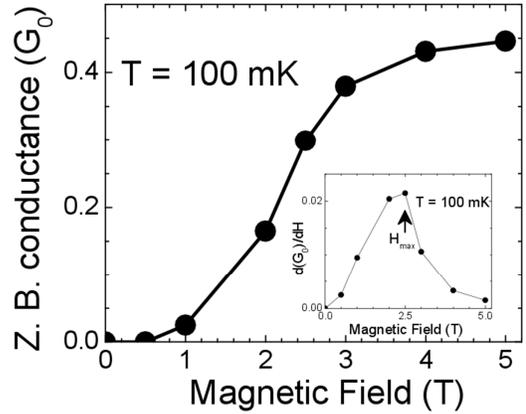}
\caption{\label{fig1} In the main panel we plot the zero bias conductance values as a function of applied magnetic field at T = 100 mK. The derivative of the curve is shown in the inset, with the peak in the derivative marked by an arrow as $H_{max}$ at 2.5 T.}
\end{figure}

Though the microscopic origin of the disorder needs substantial studies of the sample properties at the nanoscale, the possible role of crystalline boundaries, the local thickness variations or the impurities from the precursors during sample preparation (chlorine impurities in the case of ALD deposited TiN films) as well as partial oxidation of the surface in the ambient atmosphere can not be ruled out. In Fig. 7, we show the surface image (in panel (a)), measured separately over an area 28.4 nm x 28.4 nm, which very close to the interface between two regions, (refer panel (b)). The scale in panel (a) shows the height variation in this area, which is within 0.5 nm for the majority of this part of the film. The atomic planes are seen simultaneously in the most of the image, and three different orientations can be identified involving crystalline boundaries, one across the central demarking region, between the two crystallites at the top and bottom on the left part of the image. Over this area, we measured 128 x 128 conductance curves, normalized as before, and the spatial conductance changes at 0.25 mV is shown in panel (b), with the color scale between 0 to 1. The two conductance behavior, namely, with and without quasi-particle peaks are seen at dark blue and red regions, while the interface region, shows slightly less developed peaks (inset in panel (b)), but more rounding of the curves near the zero bias voltage. Across the boundary between well connected crystallites on the top left and bottom left, the local electronic properties shows uniform behavior, however, across the not so well connected crystallites, i.e., crossing over to the central shallow region, the local electronic properties vary to an extent that the quasi-particle peaks are absent in the conductance curves. The interface region, highlighted using light blue color in panel (b), corresponds to an intermediate behavior in the conductance curves as shown in the inset. The conductance in the intermediate curve remains comparatively lower in the range from 0.35 mV to 0.45 mV and possibly produces the white lines, which were seen in the bias voltage dependence in Fig. 4.

\begin{figure}[h]
\includegraphics[width=0.48\textwidth]{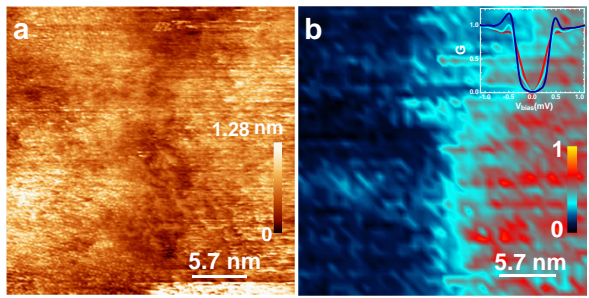}
\caption{\label{fig1} In the panel (a) we show the topographic image at 100 mK over an area 28.4 nm x 28.4 nm. The verticle scale on the right of the image shows the height variations in this area. In panel (b), the colored conductance map of 128 x 128 values, in the normalized scale from 0 to 1 is shown. The representative conductance curves in each region is shown with the same color in the inset in panel (b).}
\end{figure}

Our zero field data imply that the superconducting conductance curves deviates from the s-wave BCS-like behavior, with reduced quasi-particle peaks. These are typical in the STM measurements close to the superconductor to insulator transition in the disordered superconducting films. Our measurement suggests that the Cooper pairing emerges out of an electronic structure with a strongly reduced density of states at the Fermi level. Remarkably, the energy range for this reduction coincides with the superconducting gap. The reduced tunneling conductance features far from BCS s-wave theory can be related to strongly reduced electronic states close to the Fermi level. We do not observe such visible spatial changes at high magnetic fields to be able to relate to the inhomogenieties in the superconducting phase. Therefore we believe the spatial changes in the local properties in the superconductins phase reflect a displacement of electronic states, and the total amount of states remains roughly the same in the normal state, at the energy range studied here.

The disordered superconducting TiN film contains regions, with loosely connected crystallites suggesting that the microscopic drive towards the disorder induced insulating phase might set in locally in some areas. Large inhomogeneities appearing at the local scale play an important role, hitherto not sufficiently discussed. Furthermore, it might be possible that the superconducting and insulating regions can co-exist if the local disorder segregates two behaviors with the interface exhibiting intermediate curves with relatevely energy independent density of states.

\end{document}